\documentclass[aps,
preprint,
superscriptaddress,flushbottom,floatfix,nobibnotes,nofootinbib,a4paper,showpacs,amsmath]{revtex4-1}
\usepackage{amssymb}
\usepackage{graphicx}

\usepackage{color}

\newcommand{\be}{\begin{equation}}
\newcommand{\ee}{\end{equation}}
\newcommand{\bea}{\begin{eqnarray}}
\newcommand{\eea}{\end{eqnarray}}

\newcommand{\nn}{\nonumber}

\def\siml{{\ \lower-1.2pt\vbox{\hbox{\rlap{$<$}\lower6pt\vbox{\hbox{$\sim$}}}}\ }}

\def\siml{{\ \lower-1.2pt\vbox{\hbox{\rlap{$<$}\lower6pt\vbox{\hbox{$\sim$}}}}\ }}

\begin{document}
\title
{Proton radius from electron-proton scattering and chiral
perturbation theory} 
\author{Marko Horbatsch}
\author{Eric A. Hessels}
\affiliation{Department of Physics and Astronomy, 
York University, Toronto, Ontario M3J 1P3, 
Canada}
\author{Antonio Pineda}
\affiliation{Department of Physics and IFAE-BIST, 
Universitat Aut\`onoma de Barcelona, 
E-08193 Bellaterra, Barcelona, Spain}
\date{\today}
\begin{abstract}
We determine the root-mean-square proton charge radius, 
$R_{\rm p}$, 
from a fit to low-$Q^2$ 
electron-proton elastic-scattering cross-section data 
with the higher moments fixed 
(within uncertainties) 
to the values predicted by chiral perturbation theory. 
We obtain 
$R_{\rm p}=0.855(11)$ fm. 
This number falls between the value obtained 
from muonic hydrogen analyses and 
the CODATA value 
(based upon atomic hydrogen spectroscopy 
and electron-proton scattering
determinations).
\end{abstract}
\pacs{12.39.Fe, 13.60.Fz, 11.10.St, 12.39.Hg, 12.20.Ds, 06.20.Jr, 13.40.Gp, 14.20.Dh, 25.30.Bf}
\maketitle

The measurement 
\cite{Pohl:2010zza,Antognini:1900ns} 
of the Lamb shift in muonic hydrogen, 
$E(2P_{3/2})-E(2S_{1/2})$,
and its associated determination 
of the root-mean-square electric charge radius of the proton,
\be
\label{rpPSI}
R_{\rm p}= 0.8409(4)\; {\rm fm,}
\ee
has led to a lot of controversy. 
The reason is that this determination is 
7.1 standard deviations  
away from the CODATA 2010 
\cite{Mohr:2012tt}
value of
$R_{\rm p}= 0.8775(51)$ fm 
(and 5.6 standard deviations
away from the updated
CODATA 2014 \cite{Mohr:2015co} value of
0.8751(61) fm).
The CODATA value is based on an average of 
determinations coming from hydrogen spectroscopy 
and from electron-proton scattering data. 

Such a large discrepancy calls for an explanation. 
For the $R_{\rm p}$ value obtained from the 
Lamb shift in muonic hydrogen, 
the major criticism concerned the 
determination and error analysis 
of the two-photon-exchange contribution
needed for obtaining $R_{\rm p}$. 
On the one hand, 
dispersion relation analyses were used 
(see, 
for instance,
\cite{Pachucki:1999zza,
Martynenko:2005rc,
Carlson:2011zd,
Birse:2012eb,
Gorchtein:2013yga}).
These assume Regge behavior at large energies, 
which, 
at present, 
cannot be directly derived from QCD. 
They also require very precise knowledge 
of the elastic and inelastic form factors 
that enter into those dispersion relations, 
as in some cases very precise cancellations 
may occur. 
Nevertheless, 
the major concern is that momentum-dependent 
subtraction functions were needed to make 
the dispersion relation integrals convergent. 
Such functions cannot be deduced from experiment 
and therefore introduce some model dependence, 
which is difficult to quantify, 
as emphasized in Ref.~\cite{Hill:2011wy}.
 
Chiral perturbation theory avoids 
all of these issues 
in the calculation of the two-photon-exchange contribution. 
This contribution is chirally divergent
(linearly in 
$1/m_{\pi}$), and 
this linear divergence 
allows for a model-independent prediction 
for the two-photon-exchange term, 
which avoids the above-mentioned 
dispersion relation analysis shortcomings. 
The two-photon-exchange contribution 
has been obtained in a series 
of papers~\cite{Pineda:2004mx,
Nevado:2007dd, Peset:2014yha,
Peset:2014jxa, 
Alarcon:2013cba}.
The complete result can be found in 
Ref.~\cite{Peset:2014jxa}, 
where, not only the strict chiral result, 
but also the leading contribution associated 
with the $\Delta$ particle 
(motivated in the large-$N_{\rm c}$ approximation of QCD), 
is incorporated.

The introduction of this result into 
the muonic hydrogen bound-state energy computation 
(which was done using effective field theory techniques, 
see 
\cite{Pineda:1997bj,
Pineda:1997ie,
Pineda:1998kn,
Jenkins:1990jv,
Caswell:1985ui}) 
produced the following value \cite{Peset:2015zga}:
\be
\label{rpEFT}
R_{\rm p}= 0.8413(15)\; {\rm fm}
\,.
\ee
This value has a larger uncertainty than 
Eq.~(\ref{rpPSI}), 
but,
nevertheless, 
it eliminates the model dependence,
giving a model-independent significance to the 
substantial discrepancy with the CODATA value.

The other side of the discrepancy comes from 
the CODATA 
average~\cite{Mohr:2012tt,
Mohr:2015co}. 
As mentioned, 
the CODATA value is an average of determinations 
coming from hydrogen spectroscopy 
and from electron-proton scattering. 
In this paper we focus on the determination 
of $R_{\rm p}$ from the precise
electron-proton scattering measurements 
of the MAMI 
collaboration~\cite{Bernauer:2010wm}.
Their full analysis~\cite{Bernauer:2013tpr} leads to 
$R_{\rm p}=0.879(8)$, 
in variance with the muonic hydrogen value by 
4.6 standard deviations. 

The determination of $R_{\rm p}$ 
from electron-proton scattering data has been discussed 
extensively in the literature~\cite{BernauerThesis,
Bernauer:2013tpr,
PhysRevLett.107.119101,
PhysRevLett.107.119102,
lorenz2012size,
sick2012problems,
pohl2013muonicJ,
adamuvsvcin2013advanced,
kraus2014polynomial,
Lorenz:2014vha,
Lorenz:2014yda,
carlson2015proton,
Lee:2015qda,
arrington2015evaluation,
pacetti2015proton,
Griffioen:2015hta,
Higinbotham:2015rja,
Horbatsch:2015qda}.
The proton radius can be determined 
\cite{Horbatsch:2015qda} from 
scattering data from
\be
R_{\rm p}=
\sqrt{-3 \frac{\rm d \sigma_{\rm red}}{{\rm d}Q^2}
\Big{|}_{Q^2=0}
+\frac{3\mu_{\rm p}^2}{4 m_{\rm p}^2}},
\label{eq:REfromGE}
\ee
with
\be
\sigma_{\rm red}=(1+\tau)\frac{{\rm d}\sigma}{{\rm d}\Omega}
\Big/ \frac{{\rm d}\sigma_{\rm Mott}}{{\rm d}\Omega}=
G_{\rm E}^2+\frac{\tau G_{\rm M}^2}{ \epsilon }  ,
\label{eq:ratioToMott}
\ee
where ${\rm d}\sigma_{\rm Mott}/{\rm d}\Omega$ 
is the Mott differential cross section,  
\be
\epsilon=\left[
1+
\frac{4Q^2+\frac{Q^4}{m_{\rm p}^2}}
{8E^2-2\frac{Q^2}{m_{\rm p}}(2E+m_{\rm p})}
\right]^{-1},
\nonumber
\ee
$\tau$=$Q^2/(4m_{\rm p}^2)$,
$E$ is the electron energy,
and
$Q^2$=$-(p_{\rm i}-p_{\rm f})^2$, 
with 
$p_{\rm i}$ and $p_{\rm f}$ 
being the initial and final 
electron
four-momenta. 
Here, 
$m_{\rm p}$ is the proton mass,
$\mu_{\rm p}$=2.7928474 is the magnetic moment 
of the proton in units of nuclear magnetons,
and we use units with $\hbar=c=1$. 
Note that Eq.~(\ref{eq:REfromGE}) follows from 
Eq.~(\ref{eq:ratioToMott}) since 
$\epsilon=1$ at $Q^2$=0 for any energy $E$.

In principle, 
$R_{\rm p}$
could be determined from Eq.~(\ref{eq:REfromGE}) 
from sufficiently-precise measurements of 
${\rm d}\sigma/{\rm d}\Omega$ at small 
$Q^2$, 
because only the leading terms of the Taylor expansions of the form factors,
\be
\label{GEfit}
G_{\rm E}(Q^2) = 1 - \frac{R_{\rm p}^2}{3!} Q^2 
+ \frac{\langle r^4 \rangle_{\rm E}}{5!} Q^4 
-  \frac{\langle r^6 \rangle_{\rm E}}{7!} Q^6
+ . . . 
\,,
\ee
and
\be
\label{GMfit}
\frac{G_{\rm M}(Q^2)}{\mu_{\rm p}} = 1 - \frac{\langle r^2 \rangle_{\rm M}}{3!} Q^2 
+ \frac{\langle r^4 \rangle_{\rm M}}{5!} Q^4
-  \frac{\langle r^6 \rangle_{\rm M}}{7!} Q^6
+ . . .
  \,,
\ee
become necessary. 
However, 
with existing data, 
a functional form for the Sachs form factors 
$G_{\rm E}(Q^2)$ and $G_{\rm M}(Q^2)$ 
must be assumed
to obtain a sufficiently-accurate extrapolation of the 
measured data to 
$Q^2=0$.
The first
derivative of $G_{\rm M}(Q^2)$ and 
the second derivative of $G_{\rm E}(Q^2)$ at $Q^2$=0 
(which are proportional to the magnetic and electric 
moments 
$\langle r^2 \rangle_{\rm M}$  
 and 
$\langle r^4 \rangle_{\rm E}$,
respectively)
are of particular importance 
in this extrapolation to $Q^2$=0. 

One can find strong arguments for why one 
should focus on the low-$Q^2$ part of the data 
in order to extract the proton charge radius.
The charged-pion production threshold at 
$Q^2=- 4 m_\pi^2 \approx -0.078 \ \rm GeV^2$ 
results in a branch cut in the analytically continued 
form factor.
Thus, 
one can seriously doubt attempts at fitting 
(by polynomials or other functions, 
such as splines) 
data beyond the value of $Q^2=0.078 \ \rm GeV^2$, 
and having confidence in an accurate determination
of the slope of $G_{\rm E}(Q^2)$ at $Q^2=0$. 
Fits that include higher-$Q^2$ MAMI data 
\cite{Bernauer:2010wm}
also require floating 31 normalization constants, 
and the floating of these constants leads to considerable
flexibility in the fits, 
which also makes determination of the higher-order moments
more difficult.

However, 
concentrating only on low-$Q^2$ data
($Q^2<0.078  \ \rm GeV^2$)
has not allowed for an accurate determination of $R_{\rm p}$, 
since this data cannot determine the necessary higher moments
(in particular, $\langle r^2 \rangle_{\rm M}$
and
$\langle r^4 \rangle_{\rm E}$)
to sufficient accuracy
to allow for a precise extrapolation to $Q^2$=0 
of the required first derivative of $G_{\rm E}(Q^2)$
(Eq.~(\ref{eq:REfromGE})).
Thus, 
out of necessity, 
many attempts have been made to 
fit scattering data up to higher $Q^2$ 
to determine $R_{\rm p}$ 
while simultaneously determining the higher-order
moments. 

In ref.~\cite{Horbatsch:2015qda}, 
it was shown that values of $R_{\rm p}$
ranging from 0.84 to 0.89 fm are possible from 
acceptable fits of the MAMI data, 
with the value of $R_{\rm p}$ 
depending on the functional forms of 
$G_{\rm E}$ and $G_{\rm M}$ that are  
used for the extrapolation to $Q^2$=0. 
In particular, 
the higher moments assumed by the 
different functional forms lead to the 
wide range of $R_{\rm p}$ values.
The implication of that work~\cite{Horbatsch:2015qda}
is that a precise value of $R_{\rm p}$ cannot be obtained from
electron-proton elastic scattering, 
unless precise lower-$Q^2$ data become available,
or unless there are external constraints on the 
functional form of $G_{\rm E}(Q^2)$ and 
$G_{\rm M}(Q^2)$. 
The latter 
(external constraints on the functional forms --
as obtained from chiral perturbation theory) 
is the main topic of this work.

\begin{table}[h]
\addtolength{\arraycolsep}{0.2cm}
$$
\begin{array}{|l||c|c|c|c|c|c|}
\hline
  & \langle r^4 \rangle_{\rm E} & \langle r^6 \rangle_{\rm E} & \langle r^8 \rangle_{\rm E} & \langle r^2 \rangle_{\rm M} & 
  \langle r^4 \rangle_{\rm M} & 
  \langle r^6 \rangle_{\rm M}
\\ \hline\hline
\pi 
& 
0.71(36)   
& 
5.4(2.7)
&
104(52) 
& 
0.35(18)
& 
0.71(35) 
&
6.3(3.2)
\\
\pi\&\Delta 
& 
0.60(29)      
& 
5.0(2.0)
& 
99(37)
&
0.44(16)
& 
0.79(28)
&
6.9(2.4)
\\ \hline
\end{array}
$$
\caption{{\it Values of 
$\langle r^n \rangle_{\rm E}$ and 
$\langle r^n \rangle_{\rm M}$ in units of {\rm fm} 
from chiral perturbation theory. 
The first 
row is the pure chiral prediction 
(with only pions),
and the second row is the result after the inclusion 
of the effects associated with the $\Delta$ particle.
Uncertainties in the last two digits are shown 
in parentheses.}}
\label{tab:rn}
\end{table}

As discussed above, 
the introduction of computations that 
incorporate the correct chiral structure and 
associated power counting of the theory 
has allowed for a solution to the theoretical 
problems that the muonic hydrogen determination 
of $R_{\rm p}$ was facing, 
allowing for a model-independent determination 
of the two-photon exchange using pure 
chiral perturbation theory. 

Here we investigate if a similar analysis can 
shed some light on 
electron-proton scattering, 
and, indeed, something similar happens here. 
Chiral perturbation theory determines the dominant 
(nonanalytic) 
dependence on the pion mass of the different moments.
For $R_{\rm p}^2$=$\langle r^2 \rangle_{\rm E}$, 
one obtains only the logarithmic dependence, 
and therefore no accurate estimate can be made.
For the higher moments, 
and all magnetic moments, 
however, 
one obtains answers for the moments that are in
inverse powers of the pion mass, 
which allows for a determination of the leading term 
and an estimate of the uncertainty based on the 
estimated effect of missing
higher terms.
To obtain these moments, 
one needs the chiral expressions of 
the Sachs form factors, first obtained in 
Refs.~\cite{Gasser:1987rb,Bernard:1992qa,Bernard:1998gv}. 
The latter two references work in the heavy-baryon 
formalism to third order,
which is the formalism that we use. 
Reference~\cite{Bernard:1998gv} 
also incorporates the $\Delta$-particle effects into the computation.
Below the threshold
imposed by the branch cut
($Q^2 < 0.078 \ \rm GeV^2$), 
these form factors are analytic,
and can be Taylor expanded. 
From the Taylor expansion
and Eqs.~(\ref{GEfit}) and (\ref{GMfit}),
the moments can be determined. 
In Ref. ~\cite{Peset:2014jxa},
one can find analytic expressions for 
$\langle r^{2k} \rangle_{\rm E}$ ($k>1$). 
The expressions
for $\langle r^{2k} \rangle_{\rm M}$ 
are given in the Appendix of the present paper, 
where we also give the purely
chiral result for the Sachs form factors in a compact form.

Values (and uncertainties) 
obtained in this way for the lowest calculable 
electric and magnetic moments can be found in 
Table~\ref{tab:rn}. 
Note that in all 
cases the correction due to the $\Delta$ particle 
is quite small. 
The uncertainties in the first row of the table are 
estimated to be of order 
$m_{\pi}/\Delta \sim 1/2$,
where $\Delta=m_{\Delta}-m_{\rm p}$.
The uncertainty in the chiral perturbation theory 
contribution in the second row is 
of order 
$m_{\pi}/(m_{\rm Roper} - m_{\rm p})  \sim 1/3$, 
due to the Roper resonance. 
This estimate is also large enough to include 
corrections of order $m_{\pi}/m_{\rho}$, 
where $m_{\rho}$ is the mass of the $\rho$ meson.
Finally, 
the uncertainty of the contribution 
due to the $\Delta$ included in the
second row is estimated to be of order  
$1/N_c$ (where $N_c$=3 is the number of colors). 
In practice, we take the more conservative estimate of 
$1/2 \sim \Delta/(m_{\rm Roper} - m_{\rm p})$.
A detailed discussion of the uncertainty analysis is given in 
Ref.~\cite{Peset:2014jxa}.
%
%
%
The overall uncertainty in the moments given in the final row 
of the table is 
approximately 30 to 40\%. 

To determine $R_{\rm p}$, 
we fit the lowest-$Q^2$ MAMI data 
using Eqs.~(\ref{GEfit}) and (\ref{GMfit})
(along with Eq.~(\ref{eq:ratioToMott})), 
but fix the moments,
$\langle r^{2k}\rangle_{\rm E}$ 
(for $k \geq 2$) and 
$\langle r^{2k}\rangle_{\rm M}$ 
(for $k \geq 1$),
to be their predicted values 
(2nd row of table I) 
from chiral perturbation theory
(to within their uncertainties), 
while performing a least-squares fit to 
determine $R_{\rm p}$. 
The value of 
$\langle r^{2}\rangle_{\rm E}$$\equiv$$R_{\rm p}^2$ 
is the only moment which chiral perturbation theory 
cannot determine with sufficient accuracy,
and thus it must be determined from fitting to 
the data.

Our fitting procedure follows that described in 
Ref.~\cite{Horbatsch:2015qda}.
Before performing the fits, 
we remove the Feshbach two-photon-exchange correction
and replace it with the more complete two-photon-exchange
correction calculated following the prescription of 
Ref.~\cite{Borisyuk:2012}. 
For the low-$Q^2$ data used in this work, 
these corrections agree 
well with the low-$Q^2$ 
two-photon-exchange calculations of
Ref.~\cite{borisyuk2007two}.
In fact, 
at the very low $Q^2$ which are of interest to the present
work, 
the replacement of the Feshbach correction is not very relevant, 
despite the fact that the Feshbach correction 
ignores magnetic effects. 

We perform fits to subsets of the MAMI data 
of variable length, 
starting from the lowest available 
$Q^2$ value up to some chosen cut-off 
$Q_{\rm max}^2$ values. 
Depending on our choice of 
$Q_{\rm max}^2$,
there are still 
either 4 or 5 normalization constants 
that must also be 
determined by the fits. 
Our fits float these normalization constants, 
along with the one remaining constant 
$R_{\rm p}$ 
(the only one that cannot be determined 
from chiral perturbation theory).
In all cases, the fits return normalization
constants near unity 
(within 0.5\% in all cases, i.e., 
well within the $2 \%$ absolute
normalization uncertainty of the measurements).

\begin{figure}[t]
\includegraphics[width=6.2in]{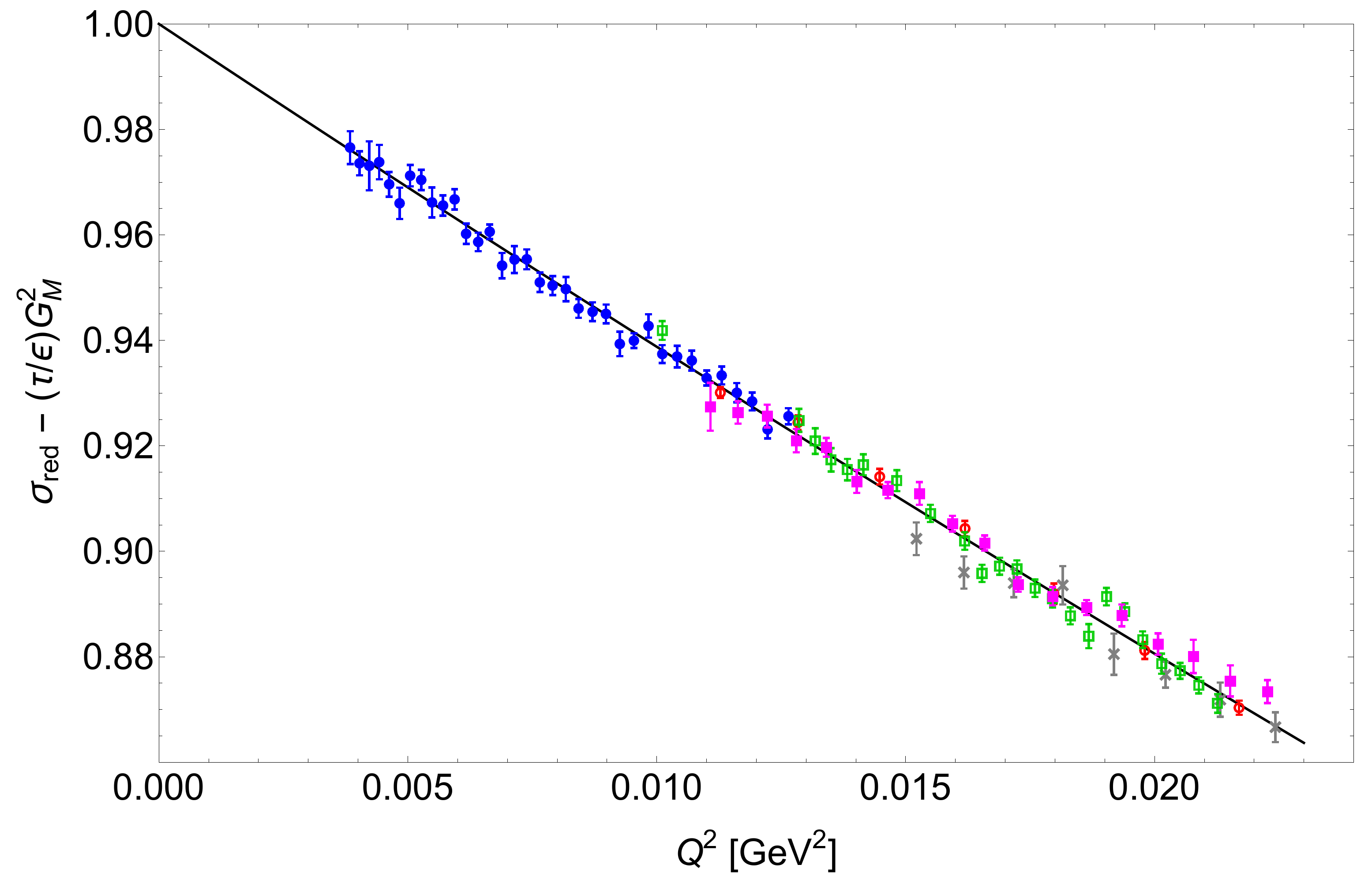}
\caption{\label{fig:fit} (Color online) 
Shown is the fit of the MAMI data 
\cite{Bernauer:2010wm} up to 
$Q_{\rm max}^2$=0.023 GeV$^2$
(the largest $Q_{\rm max}^2$ considered 
in this work). 
The quantity plotted is the experimental cross section,
scaled as in Eq.~(\ref{eq:ratioToMott}), 
with the 
contribution from the magnetic form factor 
(as predicted from chiral perturbation theory) 
subtracted out. 
The derivative of this quantity with respect to 
$Q^2$ at $Q^2$=0 gives the value of $R_{\rm p}$ 
(Eq.~(\ref{eq:REfromGE})). 
The fit uses the functional forms of Eqs.~(\ref{GEfit}) 
and (\ref{GMfit}),
with $R_{\rm p}$ floating and the higher moments 
set to the values of Table~\ref{tab:rn}.
The different symbols (colors) represent separate data groups, 
with the closed circles (blue), open circles (red), and open squares (green) data taken at an
energy of 180 MeV, 
the closed squares (magenta)
at 315 MeV,
and the crosses (gray) at 450 MeV.
Each group has a separate normalization constant
that must also float in the fit. 
Repeated measurements within a data group 
at identical (or nearly identical) $Q^2$ have been
averaged only for the purpose of aiding the 
clarity of the presentation. 
}
\end{figure}

Fig.~\ref{fig:fit} shows a typical fit used in this work. 
Shown in that plot are the experimental MAMI measurements
(scaled to give $\sigma_{\rm red}$, 
as in Eq.~(\ref{eq:ratioToMott})), 
with the magnetic form factor contribution 
$(\tau/\epsilon) G_{\rm M}^2$
(as calculated from chiral perturbation theory --
Eq.~(\ref{GMfit}) and Table~\ref{tab:rn}) 
subtracted out. 
This difference is an estimate of $G_{\rm E}^2$,
and the derivative of this quantity with respect to 
$Q^2$ at $Q^2$=0 gives the value of $R_{\rm p}$,
as indicated in Eq.~(\ref{eq:REfromGE}). 
The data shown in the figure represent 270 
measured cross sections, within five data groups 
(shown by separate symbols), 
with each data group
having a separate floating normalization constant. 
The fits use the original 270 MAMI measurements and 
their uncertainties. 
For clarity of presentation, the figure shows the average
of cross sections taken at identical 
(or nearly identical)
$Q^2$.
The fits are performed with the central 
value predicted for  
$\langle r^{2k}\rangle_{\rm E}$ 
(for $k \geq 2$) and 
$\langle r^{2k}\rangle_{\rm M}$ 
(for $k \geq 1$), as
shown in the final row of Table~\ref{tab:rn}.
The fits are then repeated for the full range of 
values for these moments that fall within the 
uncertainties given in Table~\ref{tab:rn}.

\begin{figure}[t]
\includegraphics[width=6.2in]{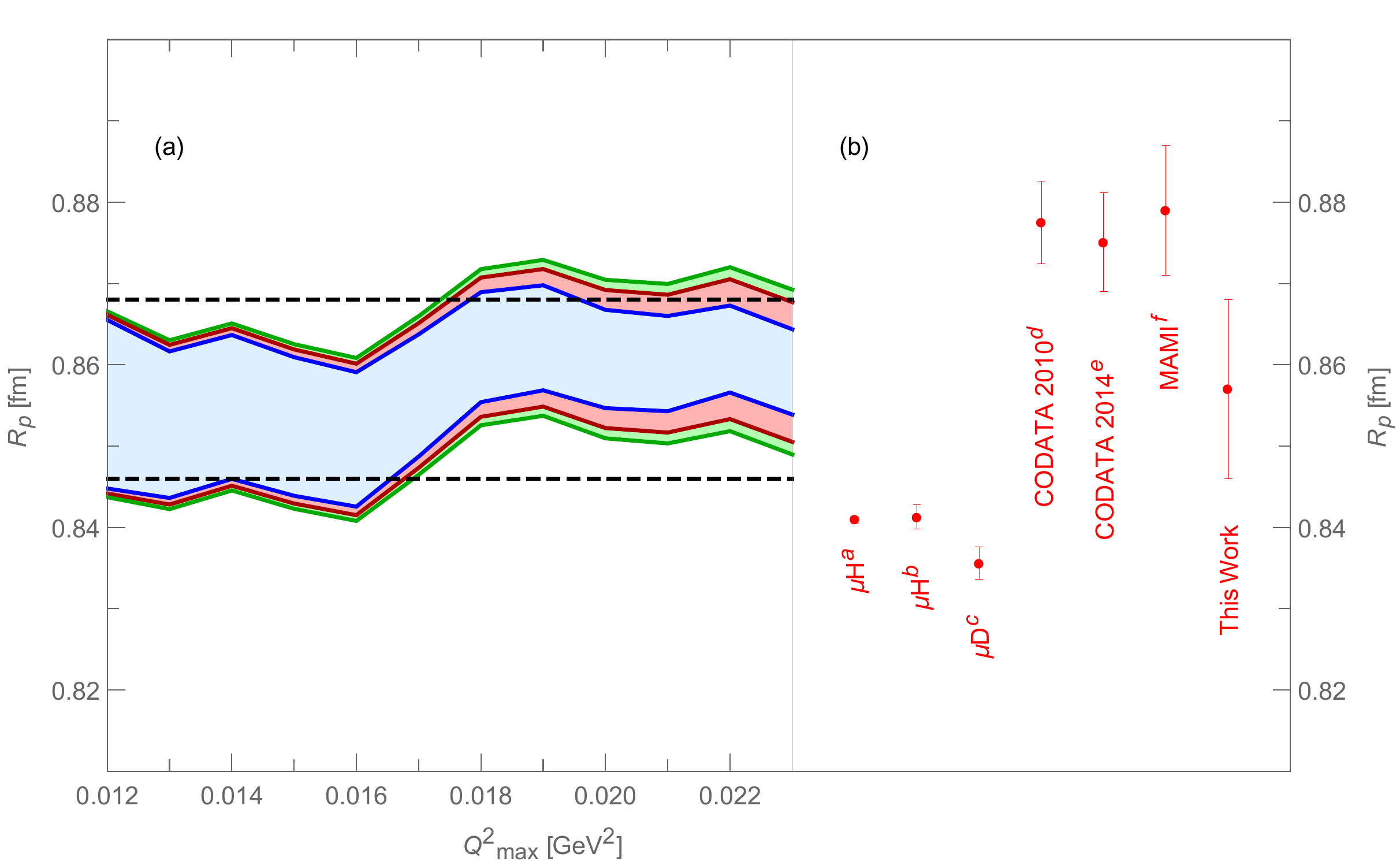}
\caption{\label{fig:bands} (Color online) 
(a) The range of $R_{\rm p}$ allowed by the fits as
a function of $Q^2_{\rm max}$. The  
central (blue) region is the range of uncertainty from the 
fits. The expanded (pink) range includes the uncertainty in 
$\langle r^2 \rangle_{\rm M}$. 
The largest (green) band
includes the uncertainties in all of the higher-order 
moments, and represents the full uncertainty in the 
determination. 
The dashed lines show the 0.855(11) fm
range of Eq.(\ref{rpUS}), which represents
the final result of this work.
(b) A comparison of the result of this work 
to the original 
(Refs.~\cite{Pohl:2010zza,Antognini:1900ns}$^a$), 
and chiral perturbation theory motivated 
(Ref.~\cite{Peset:2015zga}$^b$)
determinations of $R_{\rm p}$ 
from muonic hydrogen~\cite{Pohl:2010zza,Antognini:1900ns},
from muonic deuterium (Ref.~\cite{muonicDeut}$^c$, 
along with the hydrogen-deuterium 1S-2S 
isotope shift of Ref.~\cite{isotopeShift}),
CODATA 2010 (Ref.~\cite{Mohr:2012tt}$^d$) 
and 2014 (Ref.~\cite{Mohr:2015co}$^e$), 
and the MAMI analysis of their electron-proton elastic
scattering data (Ref.~\cite{Bernauer:2013tpr}$^f$).
}
\end{figure}

The results of the radius $R_{\rm p}$ from these fits are shown in Fig.~\ref{fig:bands}(a), 
as a function of $Q^2_{\rm max}$, the maximum $Q^2$
data that is included in the fit. 
The range of $Q^2_{\rm max}$
used is limited at the lower end by the inability to obtain a precise
fit using the small amount of MAMI data with very low $Q^2$. 
At the upper end it is limited by three concerns. 
First, we wish to stop well before 
$Q^2=0.078 \ \rm GeV^2$, 
where, 
due to the charged-pion threshold
(at $Q^2=-0.078$ GeV$^2$), 
the Taylor series is no longer meaningful.
Second, we restrict ourselves to $Q^2_{\rm max}$ values that 
give fits with a reduced $\chi^2$ of unity or better. 
Third, we are restricted by the fact that the uncertainty in 
the chiral perturbation theory predictions for the 
$\langle r^2 \rangle_{\rm M}$ and higher moments make 
a precise determination of $R_{\rm p}$ infeasible at larger
$Q^2_{\rm max}$ values. The sensitivity to 
$\langle r^2 \rangle_{\rm M}$ becomes 
particularly acute for large scattering angles, which
have a small value of $\epsilon$ and therefore 
a large $(\tau/\epsilon) G_{\rm M}^2$ contribution 
for the 180-MeV scattering with $Q^2$ above our range of
$Q^2_{\rm max}$.

The uncertainties in the determination of $R_{\rm p}$ 
are due to a combination of the statistical error from 
the least-squares fit 
(the central (blue) region in Fig.~\ref{fig:bands}(a)),
and the uncertainties in the moments 
(the largest from
$\langle r^2 \rangle_{\rm M}$, 
but also from higher electric 
and magnetic moments).  
Note that the uncertainty in the determined 
$R_{\rm p}$ at the left of Fig.~\ref{fig:bands}(a)
($Q^2_{\rm max}$=0.012 GeV$^2$, 
which includes 118 MAMI cross sections) 
is dominated by the statistical uncertainty associated
with the least-squares fit. 
The value here is 
\be
R_{\rm p}=
0.8538(104)_{\rm f}(42)_{\rm M2}(35)_{\rm E4}(1)_{\rm M4}(2)_{\rm E6}\ {\rm fm} = 0.8538(117) \ {\rm fm}.
\label{rpLOW}
\ee
Here the f subscript indicates the fit uncertainty, 
and the M2, E4, M4, E6 subscripts indicate the uncertainties
that result from the uncertainties in the moments of 
Table~\ref{tab:rn} (in the second row).  
The uncertainty at the right 
side of Fig.~\ref{fig:bands}(a)  
($Q^2_{\rm max}$=0.023 GeV$^2$,
which is a fit of 270 measured MAMI cross sections)
is dominated by the Table~\ref{tab:rn}
uncertainties:
\be
 R_{\rm p} = 
  0.8566(53)_{\rm f}(78)_{\rm M2}(59)_{\rm E4}(4)_{\rm M4}(6)_{\rm E6} \ {\rm fm}
= 0.8566(112) \ {\rm fm}. 
\label{rpHIGH}
\ee
The fact that the uncertainties at the two ends are dominated 
by different concerns, and the fact that the values agree at both ends,
adds strength to our determination. 
We average the values at the two ends to obtain our 
final determination of $R_{\rm p}$:
\be
R_{\rm p}=0.855(11)\  {\rm fm},
\label{rpUS}
\ee
where the uncertainty is chosen to be consistent 
with the whole range of $Q^2_{\rm max}$ 
shown in Fig.~\ref{fig:bands}(a). 
This range is shown by the dashed lines in 
Fig.~\ref{fig:bands}(a). 

The value of $\langle r^2 \rangle_{\rm M}$ plays 
a major role in our determination, 
as can be seen from the uncertainties labeled M2
in Eqs.~(\ref{rpLOW}) and (\ref{rpHIGH}). 
Other methods for determining 
$\langle r^2 \rangle_{\rm M}$ report
\cite{
kubis2001low,
PhysRevC.86.0652063,
PhysRevC.75.035202,
Hoferichter2016,
Bernauer:2010wm}
larger values than the chiral perturbation 
theory predictions shown in Table~\ref{tab:rn}.
The Particle Data Group reports
\cite{pdg2016}
a value of 0.602(38) fm$^2$ for 
$\langle r^2 \rangle_{\rm M}$,
which is larger  
(but within the uncertainty limits) of our 
value of 0.44(16) fm$^2$ in Table~\ref{tab:rn}.
Introducing a larger value of 
$\langle r^2 \rangle_{\rm M}$ 
into our fit would decrease 
the value of $R_{\rm p}$, 
bringing it significantly closer to the value obtained 
from muonic hydrogen. 

Our final result is compared to  
other determinations of 
$R_{\rm p}$ in Fig.~\ref{fig:bands}(b). 
Our value is higher
(by 1.2 standard deviations) 
than the
values obtained from muonic hydrogen. 
It is also higher 
(by 1.7 standard deviations)
than the value that can be determined
from a combination of a muonic deuterium 
\cite{muonicDeut}
measurement along with a hydrogen-deuterium
isotope shift measurement 
\cite{isotopeShift}.
On the other hand, it is lower than the CODATA value 
by 1.6 standard deviations and the MAMI prediction
by 1.7 standard deviations.

The uncertainty in our determination of $R_{\rm p}$ 
(Eq.~(\ref{rpUS})) is an order of magnitude larger 
than that of Eq.~(\ref{rpEFT}). 
One side effect of this fact is that we do not 
have to concern ourselves with a 
possible different definition for the proton radius 
(see the discussion in Ref.~\cite{Pineda:2004mx}), 
because the difference is 
smaller than the precision we
have obtained here.   

In summary, 
we have used chiral perturbation theory inputs, 
along with  
fits of the precise MAMI electron-proton elastic 
scattering cross sections, 
to determine
the root-mean-square charge radius of the proton:
$R_{\rm p}$ = 0.855(11) fm.
This result falls between the determinations
of $R_{\rm p}$ obtained from muonic hydrogen 
(Eqs.~(\ref{rpPSI}) and (\ref{rpEFT})) and 
the CODATA value.
This work is a step on the path to resolving the proton
radius puzzle.
The work presented here should be directly 
applicable to the analysis of ongoing and planned measurements
\cite{pRad}
of elastic scattering cross sections at very low $Q^2$.

\begin{acknowledgments}
This work was supported in part by the Spanish 
grants FPA2014-55613-P, FPA2013-
43425-P, and the Catalan grant SGR2014-1450 and by the
NSERC and CRC of Canada.
\end{acknowledgments}

\bigskip

\appendix{\bf APPENDIX}

The electric and magnetic Sachs form factors can be written in the following form in the chiral limit ($x^2=Q^2/m_{\pi}^2$):
\bea
G_{\rm E}(Q^2)&=&1 - \frac{R_{\rm p}^2}{3!} Q^2 +\frac{m_{\pi}^2}{288\pi^2F_{\pi}^2 }
\{3 \sqrt{\frac{4}{x^2}+1}
\left[
4+x^2+g_A^2(8+5 x^2)
\right]
{\rm ArcCsch}[\frac{2}{x}]
\\
\nn
&&
-12-4x^2-g_A^2(24+17x^2)
\}
\\
G_{\rm M}(Q^2)&=& \mu_{\rm p} -\frac{g_A^2 m_{\rm p} m_{\pi}}{32\pi  F_{\pi}^2 }
 \left(-2+\left(\frac{4}{x}+x\right){\rm ArcTan}[\frac{x}{2}]\right)
 \eea

For $k \geq 1$ we have the following expression for the 
magnetic moments 
(which, 
in addition to the pure chiral result, 
also incorporates the corrections associated 
with the $\Delta$ particle)
\bea
&&
\langle r^{2k} \rangle_{\rm M}
=
\frac{(-1)^k (2 k+1)!}{2\mu_{\rm p}}\frac{1}{m_{\pi}^{2k-1}}\frac{m_{\rm p} }{\pi  F_{\pi}^2 }
{\Bigg \{} -\frac{g_A^2 (-1)^k   }{4^{k+1} \left(1-4 k^2\right)}
\\
\nn
&&
+
\frac{g^2_{\pi N \Delta }  z^{-1} }{9\pi}
\Bigg[
\frac{(-1)^{k+1}   \Gamma^2 (k+1)}{  k \Gamma (2 k+2)}
+\frac{2}{4^{k}(1-4 k^2)} \sqrt{1-z^2} \left(\frac{z^2}{1-z^2}\right)^k \ln \left(\frac{2}{z}\right)
\\
&&
\nn
+
(-1)^k  \sqrt{1-z^2} \frac{\Gamma^2(k+1)}{\Gamma (2 k+2)}
\sum _{s=0}^{k-1 }\frac{1}{k-s}
\left(
\begin{array}{c}
 \frac{1}{2} \\
 s
\end{array}
\right)\left(\frac{z^2}{1-z^2}\right)^s 
\\
\nn
&&
- 
(-1)^k  \sqrt{1-z^2} \left(
\begin{array}{c}
 \frac{1}{2} \\
 k
\end{array}
\right)\left(\frac{z^2}{1-z^2}\right)^k \frac{\Gamma^2(k+1)}{\Gamma (2 k+2)}
\sum _{s=1}^{\infty } \frac{(2 s)!   z^{2 s}
   {}_2F_1\left(-k,-s;\frac{3}{2}-k;1-\frac{1}{z^2}\right)}{2^{2 s}s (s!)^2}
\Bigg]  
{\Bigg \}},
 \eea
where $z=m_{\pi}/\Delta$, 
$\Delta=m_{\Delta}-m_{\rm p}$,
$_2F_1$ is the hypergeometric function, 
and $\Gamma(n)$ is the Euler $\Gamma$ function. 
The numerical values for the masses and 
coupling constants are taken from 
Ref. \cite{Bernard:1998gv}, 
except for $g_A=1.2723(23)$, 
which we take from 
Ref. \cite{pdg2016}, 
and $g_{\pi N \Delta}=3/(2\sqrt{2})g_A=1.35$, 
which we have fixed to the large-$N_c$ prediction.

Note that the $z^{-1}$ terms cancel in the total sum and the general structure of the moments is the following:
\be
\langle r^{2k} \rangle_{\rm M} \sim \frac{1}{m_{\pi}^{2k-1}}
\left[1+{\cal O}\left(\frac{m_{\pi}}{\Delta}\right) \right]
\ee 
up to single logarithms, as it should.

\end{document}